\documentclass{article}
\usepackage{amsmath,amsfonts,url,bm}

 \twocolumn
\makeatletter
\long\def\proofbox#1{\gdef\@proofbox{#1}}
\proofbox{\small{\tt\@ifundefined{sourcepath}{}{\sourcepath/}\jobname}\\%
\ifx\UNDEF\Version\else\quad v.\Version, \fi\today}

\def\proofref#1{\proofbox{\small{\tt#1\par
	[edited by Tom Toffoli for personal use]\par
	{\tt\sourcepath/\jobname}, 
	\number\month/\number\day/\number\year
	\par}}}

 \def\affil#1{\\{\small\sl#1\par}}
 \long\def\author#1{\gdef\@author{#1}}
 \author{Tommaso Toffoli ({\tt tt\char"40bu.edu})\affil{Electrical and
Computer Engineering, Boston University, MA 02215}}

 \long\def\abstract#1{\gdef\@abstract{#1}}
 \abstract{}

\long\def\@firstoftwo#1#2{#1}
\long\def\@secondoftwo#1#2{#2}
\def\@ifundefined#1{%
  \expandafter\ifx\csname#1\endcsname\relax
    \expandafter\@firstoftwo
  \else
    \expandafter\@secondoftwo
  \fi}
 \@ifundefined{leftfrac}{\def\leftfrac{.32}}{}
 \@ifundefined{ritefrac}{\def\ritefrac{.68}}{}

\def\@maketitle{\newpage\noindent\leavevmode
  \begin{minipage}[t]{\leftfrac\textwidth}
    \hrule height0pt
    \@proofbox
  \end{minipage}\hfil
 \begin{minipage}[t]{\ritefrac\textwidth}
    \hrule height0pt
    \raggedleft
    \LARGE\@title\par
    \vskip4pt
    \large\@author
  \end{minipage}
  \vskip8pt
  \ifx\@abstract\@empty\else{\vskip.5em\leftskip1.25in\parskip4pt\small\@abstract\par\vskip.5em}\fi
  \noindent
  \rule{\textwidth}{0.4pt}
  \vskip16pt}
\makeatother

\makeatletter
\renewenvironment{thebibliography}[1]
     {\small \section*{\refname}%
      \list{\@biblabel{\@arabic\c@enumiv}}%
           {\settowidth\labelwidth{\@biblabel{#1}}%
            \leftmargin\labelwidth
            \advance\leftmargin\labelsep
            \usecounter{enumiv}%
            \let\p@enumiv\@empty
            \renewcommand\theenumiv{\@arabic\c@enumiv}}%
      \sloppy
      \clubpenalty4000
      \@clubpenalty \clubpenalty
      \widowpenalty4000%
      \sfcode`\.\@m}
     {\def\@noitemerr
       {\@latex@warning{Empty `thebibliography' environment}}%
      \endlist\par}
\makeatother

 \newcommand{\eg}{e.g.,}
 \newcommand{\ie}{i.e.,}
 \newcommand{\cf}{cf.}

 \makeatletter

 \DeclareRobustCommand\em
        {\@nomath\em \ifdim \fontdimen\@ne\font >\z@
                       \upshape \else \slshape \fi}

 \def\@empty{}
 \newcommand{\sectlabel}[1]{\label{sect:#1}}
 \newcommand{\eqlabel}[1]{\label{eq:#1}}
 \newcommand{\figlabel}[1]{\label{fig:#1}}
 \newcommand{\Sect}[2][]{\def\t@mp{#1}%
\section{#2} \ifx\t@mp\@empty\else\sectlabel{#1}\fi}
 \newcommand{\Eq}[2][]{\def\t@mp{#1}%
\begin{equation}#2\ifx\t@mp\@empty\notag\else\eqlabel{#1}\fi\end{equation}}
 \newcommand{\eq}[1]{(\ref{eq:#1})} % Ref. to equation to appear as, say, (2.37)
 \newcommand{\Eqaligned}[2][]{\def\t@mp{#1}%
\begin{equation}\begin{aligned}#2\end{aligned}
\ifx\t@mp\@empty\notag\else\eqlabel{#1}\fi
\end{equation}}
 \newcommand{\Eqmultline}[2][]{\def\t@mp{#1}%
\begin{multline}#2\ifx\t@mp\@empty\notag\else\eqlabel{#1}\fi
\end{multline}}
 \newcommand{\Eqgathered}[2][]{\def\t@mp{#1}%
\begin{equation}\begin{gathered}#2\end{gathered}
\ifx\t@mp\@empty\notag\else\eqlabel{#1}\fi
\end{equation}}

\newcommand{\Fig}[3][]{% [label], picture, caption
\begin{figure}[!htb]
 \centering{\leavevmode#2}%
 \caption{#3}
 \figlabel{#1}
\end{figure}                 }
 \makeatother

 \newcommand{\fig}[1]{Fig.\,\ref{fig:#1}}

 \def\cstrip#1{\setbox0=\hbox{$#1$}\kern-.5\wd0\lower2pt\box0}

 \def\Tr{\mathop{\mathrm{Tr}}\nolimits}
 \def\bra#1{\left\langle#1\right|}
 \def\ket#1{\left|#1\right\rangle}

 \let\RHO\rho
 \def\rho{\bm{\RHO}}
 \def\U{\text{\bf U}}
 \def\I{\text{\bf I}}

\begin{document}

 % \classification{03.67.-a, 05.30.-d.}
 % \keywords{Information and work; entropy defect; classical and quantum correlations.}
 \title{Heat-to-work conversion by exploiting\\full or partial correlations of
quantum particles}
 % \author{Lev B Levitin}{address={Electrical and Computer Engineering, Boston University}}
 % \author{Tommaso Toffoli}{address={Electrical and Computer Engineering, Boston University}}

 \proofbox{\small {\tt levitin/arxiv/\jobname}\\\today}
\def\affil#1{\\{\small\sl#1\par}}
 \makeatletter \long\def\author#1{\gdef\@author{#1}}\makeatother

 \author{Lev B Levitin and Tommaso Toffoli
 {\small(\url{levitin@bu.edu}, \url{tt@bu.edu})}%
\affil{Electrical and Computer Engineering, Boston University, MA 02215}}

 \abstract{It is shown how information contained in the pairwise correlations
(in general, partial) between atoms of a gas can be used to completely convert
heat taken from a thermostat into mechanical work in a process of relaxation of
the system to its thermal equilibrium state. Both classical correlations and
quantum correlations (entanglement) are considered. The amount of heat
converted into work is proportional to the entropy defect of the initial state
of the system. For fully correlated particles, in the case of entanglement the
amount of work obtained per particle is twice as large as in the case of
classical correlations. However, in the case of entanglement, the amount of
work does not depend on the degree of correlation, in contrast to the case of
classical correlations. The results explicitly demonstrate the equivalence
relation between information and work for the case of two-particle
correlations.}

\maketitle

The connection between work and information was first pointed out by
Szilard\cite{szilard29} in his analysis of Maxwell's demon, and later by
Landauer\cite{landauer61} and Bennett\cite{bennett82}. Using the setup of a
``thought experiment'' related to Gibbs's paradox, it was
shown\cite{levitin78,levitin87,levitin93} that using a system of two gases of
$N$ molecules each, with the molecules being in different (in general, nonorthogonal) quantum states $\rho^{(1)}$ and $\rho^{(2)}$, it is
possible to convert completely into mechanical work an amount of heat, taken
from a thermostat, which is proportional to the
\emph{entropy defect}\cite{levitin69} of the system
 \Eq{
        W = 2NkTI_0,
 }
 \noindent where $W$ is work, $T$ the temperature of the thermostat, and $I_0$
the entropy defect (also called ``quantum information'' or ``Levitin--Holevo
bound''\cite{bennett98,peres93}) of the system per molecule, which, in this
particular case, is
 \Eq{
        I_0 = -\Tr\rho\ln\rho
              +\frac12\sum_{i=1}^2\Tr\rho^{(i)}\ln\rho^{(i)},
        \ \ \rho = \frac12(\rho^{(1)}+ \rho^{(2)}).
 }
 \noindent Entropy defect has the meaning of information (about the microstate
of the system) associated with the selection of a subensemble (in this example,
$\rho^{(1)}$ or $\rho^{(2)}$) from the total ensemble $\rho$.

It was conjectured\cite{levitin78,levitin87,levitin93} that there exists a
general equivalence relation between information and work; namely, that by
having any information $J$ about the state of a physical system, it is
possible, by allowing the system to relax to its maximum-entropy state, to
convert into mechanical work an amount of heat $W=kTJ$ without any entropy
increase in the environment. (Of course, all information $J$ is lost in the
process, since the system reaches the thermal equilibrium state that has the
same energy and volume as the initial state.)

Since then, a number of papers (in particular,
\cite{alicki04,dahlsten09,feldman03,horodecki02,lloyd97,oppenheim02,scully01,zureck03})
has been devoted to various aspects of the connection between information and
work. However, the general equivalence relation remains up to now
unproven. (Part of the problem, in our opinion, is the still existing fuzziness
of distinction between ``mechanical work'' and ``heat.'') Moreover, all
real-life examples of getting work from a non-equilibrium system (\eg\ chemical
reactions, separated electrical charges, compressed gas) by allowing it to
relax to equilibrium are based on the information contained in one-particle
distributions---as if all the particles comprising the system were independent.
 % as if all the particles comprising the system were independent.
 To the best of our knowledge, information stored in multi-particle
distributions has never been used to extract mechanical work.

Here we address the problem of using information contained in two-particle
correlations for converting heat into work. Our analysis is based on two
important assumptions:
 \begin{enumerate}
 \item Since a unitary transformation of the system's state
does not change its entropy, it can be performed without any energy
dissipation; and 
 \item Since orthogonal quantum states are perfectly
distinguishable, there exist partitions which are permeable for one of such
states but not for the other (a well-known example of such ``partitions'' is a
light polarizer).
 \end{enumerate}

In order to avoid getting involved in definitional arguments---such as appear
sometimes in the physics literature---concerning the distinction between work
and heat, we shall consider an isothermal quasistatic process in which heat is
taken from a thermostat and eventually transferred to another system in a form
that is inequivocally mechanical work---namely, the lifting of a weight.

We shall treat in separate sections the case when the two particles are
classically correlated and that dealing with quantum correlations, and conclude
with a discussion of the overall results.

\Sect[classical]{The case of classical correlations}

Consider a gas of molecules that consist of two different atoms, $A$ and
$B$. To each atom is associated a 2-dimensional Hilbert space of states with
basis vectors $\ket0$ and $\ket1$ (a \emph{qubit}\cite{schumacher95}). The
thermal equilibrium (maximum-entropy) state of a gas molecule is described by a
density matrix
 %{\small
 \Eq{
 \rho  =\frac12(\rho_1+\rho_2),\ \text{with}\ %\text{where}\
 \begin{aligned}
  \rho_1 &=\tfrac12(\ket{00}\bra{00}+\ket{11}\bra{11}),\\
  \rho_2 &=\tfrac12(\ket{01}\bra{01}+\ket{10}\bra{10}).
 \end{aligned}
 }
 %}%end small
 \noindent States $\rho_1$ and $\rho_2$ are \emph{orthogonal}, and correspond
to (classically) \emph{maximally correlated} atoms $A$ and $B$. Each of these
states has an entropy defect (in nats) $I_\text c=\ln2$ per pair, or
$J_\text c=N\ln2$ for a gas of $N$ molecules. Note that the one-particle
probabilities of states $\ket0$ and $\ket1$ of both atoms are the same in
states $\rho_1$ and $\rho_2$ as at equilibrium. Therefore, it is impossible to
distinguish between this state and the equilibrium state on the basis of
one-particle measurements.

Without loss of generality, assume we know that the gas is in state $\rho_1$
(let us call it gas 1). Let the gas occupy a vessel of volume $2V$ divided by a
partition into two equal parts of volume $V$, each occupied by $\frac N2$
molecules, and being in thermal contact with a thermostat at temperature
$T$. To the gas on the right side of the partition apply a unitary
transformation $\U_1=\I_A\otimes \U_B$, where $\I_A$ is the identity operator
in the Hilbert space of $A$ and $\U_B$ the operator that interchanges states
$\ket0$ and $\ket1$ of $B$---that is, the Pauli matrix
$\sigma_x$. Transformation $\U_1$ converts gas 1 into a gas with density matrix
$\rho_2$ (gas 2).

Now, replace the partition by two movable semipermeable partitions such that
the partition that faces gas 1 is permeable to gas 1 but not to gas 2, and vice
versa for the other partition.  Since the total pressure of the mixed gases is
larger than the pressure of one of the gases, the two partitions will be pushed
apart---and we can use the setup shown in \fig{1} to lift weights $m_1$ and
$m_2$.

 \Fig[1]{\def\XMAG{.8}\unitlength\XMAG bp\footnotesize
 \def\HSIZ{256}\def\VSIZ{56}\def\XOFS{-128}\def\YOFS{-28}
 \begin{picture}(\HSIZ,\VSIZ)(\XOFS,\YOFS)
 \thicklines\put(-64,-18){\framebox(128,38){}}
% \thinlines \put(-4,-19){\dashbox{2}(8,40){}}
 \thinlines \put(-5,-19){\dashbox{2}(10,40){}}
 \thicklines\put(-96,-4){\circle{8}}\put(-96,-4){\makebox(0,0){.}}
            \put( 96, 0){\circle{8}}\put( 96, 0){\makebox(0,0){.}}
 \thinlines \put(-96, 0){\line(1,0){101}}\put(96,4){\line(-1,0){101}}  
            \put(-100,-4){\line(0,-1){16}}\put(100,0){\line(0,-1){20}}
            \put(-100,-25){\circle*{10}}\put(100,-25){\circle*{10}}
            \put(-36,11){\cstrip{\text{gas 1}}}
            \put( 36,11){\cstrip{\text{gas 2}}}
            \put(-36,-8){\cstrip{\rho_1}}
            \put(  0,-8){\cstrip{\rho}}
            \put( 36,-8){\cstrip{\rho_2}}

            \put(-84,-25){\cstrip{m_1}}
            \put( 84,-25){\cstrip{m_2}}
 \end{picture}
 }
% {Setup for obtaining work by mixing gases with correlated atoms in molecules.}
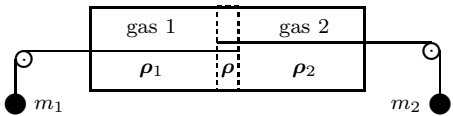
 {Scheme for obtaining work by mixing gases. The excess pressure of the gases
between the semipermeable partitions pushes them apart and lifts weights $m_1$
and $m_2$.}

Consider the moment when the right partition has advanced to the right by a
volume $V_1$ and the left partition to the left by a volume $V_2$. Gas 1 will
then fill the volume $V+V_1$ and gas 2 the volume $V+V_2$. Let us assume that
the gases are ideal and rarefied. Then the pressures of the gases to the left,
to the right, and in between the partitions are, respectively,
 \Eq{
 \begin{gathered}
 P_1 = \frac{(N/2)kT}{V+V_1},\ P_2 = \frac{(N/2)kT}{V+V_2},\\
 P_\text m= \frac{NkT}2 \Big(\frac 1{V+V_1} + \frac 1{V+V_2}\Big).
 \end{gathered}
 }
 \noindent The work produced by the gas by lifting masses of appropriate
weights in this quasistatic isothermal process  is
 \Eq{
% W = \int_0^V(P_m-P_1)dV_2+\int_0^V(P_m-P_2)dV_1=NkT\ln2.
 W_\text c = \int_0^V(P_\text m-P_2)dV_1+\int_0^V(P_\text m-P_1)dV_2=NkT\ln2.
 }
 \noindent Note that the total energy of the system has not changed in the
process. Thus, the amount of heat converted into mechanical work is
proportional to the entropy defect of the system,
 \Eq[cdefect]{
 W_\text c = NkTI_\text c = kTJ_\text c.
 }
 \noindent By the end of the process the entire vessel is occupied by a mixture
of gas 1 and gas 2 with density matrix $\rho$. The initial information about
the location of each pair of atoms, contained in the correlation of states of
the two atoms in a molecule, is now erased, and this increase in entropy
exactly compensates for the decrease of entropy of the thermostat.

\medskip

Consider now the case of partial classical correlations, when each molecule is
in a separable state
 \Eq{
 \rho_\text{1p} = \tfrac p2 (\ket{00}\bra{00}+\ket{11}\bra{11})+
          \tfrac{1-p}2(\ket{01}\bra{01}+\ket{10}\bra{10}),
 }
 where $\frac12<p\leq1$. (If $p=\frac12$, the system is in the
maximum-entropy---\ie\ equilibrium---state, and no correlation exists.) The
complementary state is
 \Eq{
 \rho_\text{2p} = \tfrac{1-p}2 (\ket{00}\bra{00}+\ket{11}\bra{11})+
                  \tfrac12 (\ket{01}\bra{01}+\ket{10}\bra{10}),
 }
 so that the equilibrium state is $\rho=\frac12(\rho_\text{1p}+\rho_\text{2p})$. Note
that the marginal probabilities for each atom to be in state $\ket0$ or $\ket1$
are equal; thus, the deviation from equilibrium is entirely due to
correlations.

The entropy defect (per molecule) of the system is now
 \Eq{
 I_\text{cp} = -\Tr\rho\ln\rho+\frac12\sum_{i=1}^2\Tr\rho_{i\text p}\ln\rho_{i\text p}
 = \ln2-h(p),
 }
 where $h(p) = -p\ln p-(1-p)\ln(1-p)$ is the binary entropy function.

Suppose the gas is in state $\rho_\text{1p}$ (gas 1p). Using the same
experimental setup as in the case of maximally correlated atoms, we apply the
unitary transformation $\U_1$ to the gas on the right side of the
partition. This transforms gas 1p into gas 2p with density matrix
$\rho_\text{2p}$.

Now let us use exactly the same movable semipermeable partitions as
before. Note that the partition that faces gas 1p performs in fact a
measurement over the state $\rho_\text{1p}$, which results either in state
$\rho_1$ (with probability $p$) or $\rho_2$ (with probability $1-p$).
Similarly, the partition that faces gas 2p produces molecules in state $\rho_2$
(with probability $p$), or in state $\rho_1$ (with probability $1-p$). As a
result, the gas between the partitions has the maximum-entropy density matrix
$\rho$.

Initially, the total pressure of the mixed gases between the partitions is
larger than the pressures of the gases to the left or to the right of both
partitions. Hence the partitions are pushed apart as in the previous case, and
the gases can produce work by lifting weights (\fig{2}). Note, however, that
the compositions of the gases beyond the partitions are changing in the
process, owing to the filtering action of the partitions (we denote changing
mixtures by gases
 \def\primetext#1{{\scriptstyle(}\text{#1}{\scriptstyle)}'}
 $\primetext{1p}$ and $\primetext{2p}$).

 \Fig[2]{\def\XMAG{.8}\unitlength\XMAG bp\footnotesize
 \def\HSIZ{256}\def\VSIZ{56}\def\XOFS{-128}\def\YOFS{-28}
 \begin{picture}(\HSIZ,\VSIZ)(\XOFS,\YOFS)
 \thicklines\put(-64,-18){\framebox(128,38){}}
% \thinlines \put(-4,-18){\dashbox{2}(8,38){}}
 \thinlines \put(-18,-19){\dashbox{2}(36,40){}}
 \thicklines\put(-96,-4){\circle{8}}\put(-96,-4){\makebox(0,0){.}}
            \put( 96, 0){\circle{8}}\put( 96, 0){\makebox(0,0){.}}
 \thinlines \put(-96, 0){\line(1,0){114}}\put(96,4){\line(-1,0){114}}  
            \put(-100,-4){\line(0,-1){16}}\put(100,0){\line(0,-1){20}}
            \put(-100,-25){\circle*{10}}\put(100,-25){\circle*{10}}
            \put(-41,11){\cstrip{\text{gas}\ \primetext{1p}}}
            \put( 41,11){\cstrip{\text{gas}\ \primetext{2p}}}
            \put(  0,12){\cstrip{\rho}}
            \put(  0,-10){\cstrip{V_1{+}V_2}}
            \put(-84,-25){\cstrip{m_1}}
            \put( 84,-25){\cstrip{m_2}}
 \end{picture}
 }
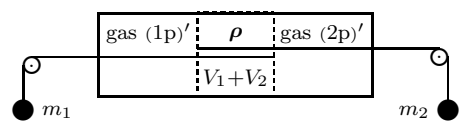
 {Scheme for converting heat to work by mixing partially correlated gases.}

As before, let the right partition be moved to the right over a volume
$V_1$, and the left partition to the left over a volume $V_2$. The
pressures of the gases to the left, the right, and in between the partitions
then become, respectively,

 \Eqaligned{
 P_\text{1p} &= \frac{NkT}2 \big(\frac p{V+V_1}+\frac{1-p}{V-V_2}\big),\\
 P_\text{2p} &= \frac{NkT}2 \big(\frac p{V+V_2}+\frac{1-p}{V-V_1}\big),\\
 P_\text{mp} &= \frac{pNkT}2 \big(\frac1{V+V_1}+\frac1{V+V_2}\big).
 }

\noindent The partitions will stop moving when the pressures on the two sides
of the partitions become equal, \ie\
 \Eq{
 P_\text{1p}=P_\text{mp}=P_\text{2p}.
 }

\noindent Solving these equations, one obtains
 \def\V{{(2p-1)V}}

 \Eq{
 V_\text{1max} = V_\text{2max} = \V.
 }

\noindent Hence, the total work produced by the gas from the heat taken from
the thermostat is

% \iftrue
 \iffalse
 \Eq[total]{
 W_\text{cp} = \int_0^{\V} \kern-32pt (P_\text{mp}-P_\text{2p})dV_1
                +\int_0{^\V}  \kern-32pt (P_\text{mp}-P_\text{1p})dV_2
                 = NkT(\ln2-h(p))=NkTI_\text{cp}.
 }
 \else
 \Eqaligned[total]{
 W_\text{cp} &= \int_0^{\V} \kern-20pt (P_\text{mp}-P_\text{2p})dV_1
                +\int_0^{\V}  \kern-20pt (P_\text{mp}-P_\text{1p})dV_2\\
             &= NkT(\ln2-h(p))=NkTI_\text{cp}.
 }
 \fi

\noindent Note that by the end of the process all three parts of the entire
vessel are occupied by gas in the equilibrium state $\rho$. Thus, the initial
information (equal to the entropy defect) is expended in the conversion of heat
into work.

\Sect[quantum]{The case of quantum correlations (entanglement)}

Consider now the case where atoms forming a molecule are in a \emph{maximally
entangled} state. There are four such states of two qubits, namely, the Bell
states\cite{bennett98} $\Phi^+$, $\Phi^-$, $\Psi^+$, and $\Psi^-$,
defined as follows
 \Eq{
 \Phi^\pm =  \frac1{\sqrt2}(\ket{00}\pm\ket{11}),\ \
 \Psi^\pm =  \frac1{\sqrt2}(\ket{01}\pm\ket{10}).
 }
 \noindent These four states are orthogonal and form a basis in the
4-dimensional tensor-product Hilbert space of those two atoms. The entropy
defect of the system in one of those states is $I_\text q=2\ln2$ per pair, or
$J_\text q=2N\ln2$ for the whole gas (as before, we assume the gas to be ideal
and the states of the molecules, independent).

\medskip

Suppose that the gas is initially in state $\Phi^+$. Then, by use of the same
transformation $\U_1$ as in the previous section, we can convert half of the
gas molecules into state $\Psi^+$. Using the same setup as in \fig{1}, we
obtain work $W_1=NkT\ln2$; the final state of the gas will be a mixture
 \Eq{
 \rho^{(1)}=\frac12(\ket{\Phi^+}\bra{\Phi^+}+\ket{\Psi^+}\bra{\Psi^+}).
 }

Now we can use a unitary transformation  $\U_2=\U_\Phi\otimes\U_\Psi$, where
$\U_\Phi$ is the unitary transformation that permutes $\Phi^+$ and $\Phi^-$ in
the subspace spanned by these orthogonal states, and $\U_\Psi$ analogously
permutes $\Psi^+$ and $\Psi^-$. Transformation $\U_2$ converts $\rho^{(1)}$
into a mixed state
 \Eq{
 \rho^{(2)}=\frac12(\ket{\Phi^-}\bra{\Phi^-}+\ket{\Psi^-}\bra{\Psi^-}).
 }
 \noindent Note that states $\rho^{(1)}$ and $\rho^{(2)}$ are orthogonal, and
thus perfectly distinguishable.

Employing transformation $\U_2$ and using the same setup as before, we can now
repeat the mixing procedure. This will yield additional work $W_2=Nk\ln2$,
bringing the system to the equilibrium state
$\rho=\frac12(\rho^{(1)}+\rho^{(2)})$. Hence, the total amount of heat
converted into work is
 \Eq[qdefect]{
 W_\text q=W_1+W_2=2NkT\ln2=NkTI_\text q=kTJ_\text q.
 }
 \noindent Since the entropy defect of the entangled states is two times as
large as that of classically correlated states, the amount of work $W_\text q$
is also larger than $W_\text c$ by a factor of 2.

\medskip

Now assume that the gas molecules are initially in a pure partially entangled
state
 \Eq{
 \psi = a\ket{00}+b\ket{01}+c\ket{10}+d\ket{11},
 }
 \noindent where
 \Eq[balance]{
 |a|^2=|d|^2=\tfrac p2,\ |b|^2 = |c|^2= \tfrac{1-p}2,\ c=b^*,\ d=-a^*.
 }
 \noindent Conditions \eq{balance} ensure that the one-particle density
matrices are the same as at equilibrium. (Entanglement vanishes when
$p=\frac12$.)  State $\psi$ displays the same partial two-particle correlation
as the mixed state $\rho_\text{1p}$. The important difference is that in this
case the correlation is entirely due to the entanglement. Thus, the
two-particle system is in a pure state and its entropy defect is $I_\text
q=2\ln2$, independently of the parameter $p$. This leads to the paradoxical
fact that the amount of work that can be obtained by the use of a gas in a
partially entangled state is the same as in the case of maximally entangled
state, namely,
 \Eq[qdefectmax]{
 W_\text q =2NkT\ln2=NkTI_\text q.
 }
 \noindent Indeed, state $\psi$ can be transformed by an appropriate unitary
operator into state $\Phi^+$, and the procedure described above can be applied
to convert to work an amount of heat equal to $W_\text q$.

\medskip \noindent
Expressions \eq{cdefect}, \eq{total}, \eq{qdefect}, and \eq{balance}
present results related to the two extreme cases of entirely ``classical'' and
entirely ``quantum'' correlations.  One could consider a more general case of a
mixed non-separable state, where the correlation is partially ``quantum''
(entanglement) and partially ``classical.''  The amount of work in this case
should fall between the values given by \eq{total} and~\eq{qdefect}.

\Sect[concl]{Conclusions}

The above results demonstrate the equivalence relation between information and
work for the case when the information is contained only in \emph{two-particle}
probability distributions, and cannot be extracted from a mere one-particle
distribution.

The situation with the equivalence between information and work in general
remains unclear and, in a way, paradoxical. On one hand, many researchers are
strongly convinced of it and express it in a very firm way, \eg ``Information
has an energetic value: It can be converted into work'' (Zureck\cite{zureck02};
but \cf\ \cite{zureck03}). (In fact, it is not \emph{information} that can be
converted into work, but \emph{heat} that can be converted into work by use of
information.)  On the other hand, in our opinion, one still lacks a convincing
general proof that \emph{any} kind of information can be equally successfully
used to convert heat into \emph{any} kind of work, so that the general
statement remains a sort of ``folk theorem.'' We believe that the root of this
problem is the fact that, in spite of many unquestionable examples of ``work''
known to physicists, there is no general rigorous definition that would
distinguish between these two forms of energy transfer---heat and work.

Informally speaking, work is an ``informed'' transfer of energy, \ie\ a
transfer such that we know exactly the change of the state of each degree of
freedom, resulting from this transfer; while heat is energy transfered in such a
way that we have no such knowledge. From that viewpoint, the equivalence
between information and work becomes indeed a tautology. However, a
rigorous formalization of these ideas has not yet been presented.

\Sect[ack]{Acknowledgments}

We wish to thank Charles Bennett for bringing to our attention a number of
related papers, and to Oscar Dahlsten and Michal Horodecki for updated
references.


\begin{thebibliography}{00}
\let\bib\bibitem
 \bib{szilard29}{\sc Szilard}, Leo, ``\"Uber die entropieverminderung in einem
thermodynamischen system bei eingriffen intelligenter wesen'' (``On the
decrease of entropy in a thermodynamic system by the intervention of
intelligent beings''), {\sl Z~Phys \bf53} (1929), 840--856.
 \bib{landauer61}{\sc Landauer}, Rolf, ``Irreversibility and heat generation in
the computing process,'' {\sl IBM J Res. Devel. \bf 5} (1961) 183--191.
 \bib{bennett82}{\sc Bennett}, Charles H, ``The thermodynamics of
computation---a review,'' {\sl Int. J Theor. Phys. \bf21} (1982), 905--940.
 \bib{levitin78}{\sc Levitin}, Lev B, ``Quantum amount of information and
maximum work,'' {\sl Proc. 13th IUPAP Conf. Stat. Phys.} ({sc D {\sc Cabile},
DG {\sc Kuper}, and I {\sc Riess}, eds.), Bristol (England), A Hilger 1978.
 \bib{levitin87}{\sc Levitin}, Lev B, ``Information theory for quantum
systems,'' in {\sl Information, Complexity, and Control in Quantum Physics},
(S {\sc Diner} and G {\sc Lochak}, eds.), Springer 1987, 15--47.
 \bib{levitin93}{\sc Levitin}, Lev B, ``Gibbs' paradox and equivalence relation
between quantum information and work,'' in {\sl Proc. Worksh. on Physics and
Computation} (PhysComp'92), IEEE Comp. Soc. 1993, 223--226.
 \bib{levitin69}{\sc Levitin}, Lev B, ``On the quantum measure of the amount of
information,'' in {\sl Proc. 4th National Conf. on Information Theory},
Tashkent, USSR, 1969, 111--116. English translation: {\sl Ann. Fond. de Broglie
\bf 21} (1996), 345--348.
 \bib{bennett98}{\sc Bennett}, Charles H, and Peter {\sc Shor}, ``Quantum
information theory,'' {\sl IEEE Trans. Info. Theory \bf44} (1998), 2724--2742.
 \bib{peres93}{\sc Peres}, Asher, {\sl Quantum Theory, Concepts and Methods},
Kluwer 1993.
 \bib{alicki04}{\sc Alicki}, Robert, Michal {\sc Horodecki}, Pawel {\sc
Horodecki}, and Ryszard {\sc Horodecki}, ``Thermodynamics of quantum
informational systems---Hamiltonian description,'' {\sl Open
Syst. Info. Dyn. \bf11} (2004), 205--217
 % arXiv:quant-ph/0402012 (2004).
 \bib{dahlsten09}{\sc Dahlsten}, Oscar, Renato {\sc Renner}, Elizabeth {\sc
Rieper}, and Vladko {\sc Vedral}, ``The work value of information,''
arXiv:quant-ph/0908.0424 (2009).
 % no journal aricle as of 101208
 \bib{feldman03}{\sc Feldmann}, Tova, and Ronnie {\sc Kosloff}, ``Quantum
four-stroke engine: Thermodynamic observables in a model with intrinsic
friction,'' {\sl Phys. Rev. E \bf68} (2003), 016101.
 \bib{horodecki02}{\sc Horodecki}, Michal, Jonathan {\sc Oppenheim}, and
Ryszard {\sc Horodecki}, ``Are the laws of entanglement theory
thermodynamical?'' {\sl Phys. Rev. Lett. \bf89} (2002), 240403.
 %arXiv:quant-ph/0207177
 \bib{lloyd97}{\sc Lloyd}, Seth, ``Quantum-mechanical Maxwell's demon,'' {\sl
Phys. Rev. A \bf56} (1997), 3374--3382.
 \bib{oppenheim02}{\sc Oppenheim}, Jonathan, Michal {\sc Horodecki}, Pawel {\sc
Horodecki}, and Ryszard {\sc Horodecki}, ``A thermodynamic approach to
quantifying quantum correlations},'' {\sl PRL \bf89} (2002), 180402.
 \bib{scully01}{\sc Scully}, Marlan, ``Extracting work from a single thermal
bath via quantum negentropy,'' {\sl Phys. Rev. Lett. \bf87} (2001), 220601.
 \bib{zureck03}{\sc Zureck}, Wojciech, ``Quantum discord and Maxwell's demons,''
{\sl Phys. Rev. A \bf67} (2003), 012320. 
 \bib{schumacher95}{\sc Schumacher}, Benjamin, ``Quantum coding,'' {\sl
Phys. Rev. A \bf51} (1995), 2738--2747,
 \bib{zureck02}{\sc Zureck}, Wojciech, ``Quantum discord and Maxwell's demons,''
arXiv:quant-ph/0202123 (2002).
\end{thebibliography}
\end{document}